\documentclass[prd,reprint,aps,amsmath,amssymb,preprintnumbers,superscriptaddress,nofootinbib]{revtex4-1}

\usepackage{subfigure}
\usepackage{enumitem}
\usepackage{color}
\usepackage{graphicx}
\usepackage{hyperref}
\usepackage{ctable}

\def\beq{\begin{equation}}
\def\eeq{\end{equation}}
\def\bea{\begin{eqnarray}} 
\def\eea{\end{eqnarray}}
\def\nn{\nonumber}

\def\eps{\varepsilon}

\newcommand{\lsim}{
\mathrel{\hbox{\rlap{\hbox{\lower4pt\hbox{$\sim$}}}\hbox{$<$}}}}
\newcommand{\gsim}{
\mathrel{\hbox{\rlap{\hbox{\lower4pt\hbox{$\sim$}}}\hbox{$>$}}}}

\newcommand{\vE}{\vec{E}}
\newcommand{\vB}{\vec{B}}

\newcommand{\vJ}{\vec{J}}

\begin{document}
    
\title{Extension of the electrodynamics in the presence of the axion and dark photon}
\author{Fa Peng Huang}
\affiliation{Center for Theoretical Physics of the Universe, IBS, Daejeon 34126, Korea}
\author{Hye-Sung Lee}
\affiliation{Department of Physics, KAIST, Daejeon 34141, Korea}

\begin{abstract}
We present the extended electrodynamics in the presence of the axion and dark photon.
We derive the extended versions of Maxwell's equations and dark Maxwell's equations (for both massive and massless dark photons) as well as the wave equations.
We discuss the implications of this extended electrodynamics including the enhanced effects in the particle conversions under the external magnetic or dark magnetic field.
We also discuss the recently reported anomaly in the redshifted 21cm spectrum using the extended electrodynamics.
\end{abstract}
\pacs{}
\maketitle

\section{Introduction}
\label{sec:introduction}
There are so many physical laws that have affected the way we understand the universe as well as the way we live our daily life.
Yet, there are not many that match the electrodynamics represented by Maxwell's equations, which were discovered in the nineteenth century.
The discovery of Maxwell's equations completely changed our daily and nightly life.

Maxwell's equations describe how the electromagnetic fields can be created with the sources (charges, currents and the time derivative of the electromagnetic fields).
In the presence of the new particles, Maxwell's equations can be effectively modified as they can play the role of the additional sources.
Interesting point is that the dark matter (more generally, dark sector particles), which are presumed not to interact electromagnetically with a sizable coupling, can be also the sources.
Although these dark sector particles may not carry any electromagnetic charges, they can still couple to the photon ($\gamma$) either through a small mixing or a suppressed nonrenormalizable operator.

In this letter, we investigate the possibility of extending Maxwell's equations in the presence of the axion and dark photon.
These particles are popular candidates of the light dark sector particles, which have been extensively studied \cite{Essig:2013lka,Alexander:2016aln}.
They can interact with the photon (or the electromagnetic fields) in the form of the portals.
The axion ($a$) is a pseudoscalar ($CP$-odd) boson, which is often motivated to solve the $CP$ problem of the QCD and the dark matter \cite{Peccei:1977hh,Weinberg:1977ma,Wilczek:1977pj,Shifman:1979if,Kim:1979if,Zhitnitsky:1980tq,Dine:1981rt}.
(We will use the term axion to indicate both the QCD axion and axion-like particle.)
It can couple to the photon through an axion portal ($a$-$\gamma$-$\gamma$ coupling).
The dark photon ($\gamma'$) \cite{ArkaniHamed:2008qn} is a vector boson, which has been studied in the context of the positron excess and the muon anomalous magnetic moment.
It can mix with the photon through a vector portal ($\gamma$-$\gamma'$ mixing) \cite{Holdom:1985ag}.

Recently, it was also pointed out that a new portal called `dark axion portal' ($a$-$\gamma$-$\gamma'$ and $a$-$\gamma'$-$\gamma'$ couplings) can open up when the axion and dark photon coexist \cite{Kaneta:2016wvf} and triggered subsequent studies in various contexts \cite{Choi:2016kke,Kaneta:2017wfh,Agrawal:2017eqm,Kitajima:2017peg,Daido:2018dmu,Choi:2018dqr,deNiverville:2018hrc}.
This new portal is completely independent and cannot be constructed with other portals.
One of the advantages we want to take in this work is that the size of this new portal can be way larger than the one constructed by the mere combination of the other two portals (vector portal and axion portal), which are highly constrained by the data.
The theoretical framework and realization for these portals were explicitly studied and documented including a new mechanism of how a dark photon can be a relic dark matter itself (for instance, see Refs.~\cite{Kaneta:2016wvf,Kaneta:2017wfh}).

We present the extended electrodynamics using the portals in the form of the extended Maxwell's equations and wave equations.
They imply many interesting phenomena.

\section{Lagrangian}
\begin{table*}[bt]
\begin{tabular}{rcl}
\hline
\\[-6pt]
\( \displaystyle \nabla\cdot \vE \)&=& \( \displaystyle \rho + G_{a\gamma\gamma}\nabla a \cdot\vB + G_{a\gamma\gamma'}\nabla a \cdot \vB' \) \\[7pt]
\( \displaystyle \nabla \times \vB \)&=&\( \displaystyle \vJ + \frac{\partial \vE}{\partial t} - G_{a\gamma\gamma}\Big(\frac{\partial a}{\partial t}\vB + \nabla a \times \vE\Big) - G_{a\gamma\gamma'}\Big(\frac{\partial a}{\partial t}\vB' + \nabla a \times \vE'\Big) \) \\[10pt]
\( \displaystyle \nabla \cdot \vB \)&=&\( \displaystyle 0 \) \\[4pt]
\( \displaystyle \nabla \times \vE \)&=&\( \displaystyle - \frac{\partial \vB}{\partial t} \) \\[9pt]
\hline
\\[-6pt]
\( \displaystyle \nabla\cdot \vE' \)&=&\( \displaystyle \big(\rho' + \varepsilon \rho\big) -m_{\gamma'}^2 A'^0 + G_{a\gamma'\gamma'}\nabla a \cdot\vB' + G_{a\gamma\gamma'}\nabla a \cdot\vB \) \\[7pt]
\( \displaystyle \nabla \times \vB' \)&=&\( \displaystyle \big(\vJ' + \varepsilon \vJ\big) - m_{\gamma'}^2\vec{A}' + \frac{\partial \vE'}{\partial t} - G_{a\gamma'\gamma'}\Big(\frac{\partial a}{\partial t}\vB' + \nabla a \times \vE'\Big) - G_{a\gamma\gamma'}\Big(\frac{\partial a}{\partial t}\vB + \nabla a \times \vE\Big) \) \\[10pt]
\( \displaystyle \nabla \cdot \vB' \)&=&\( \displaystyle 0 \) \\[4pt]
\( \displaystyle \nabla \times \vE' \)&=&\( \displaystyle - \frac{\partial \vB'}{\partial t} \) \\[9pt]
\hline
\end{tabular}
\caption{Extended version of Maxwell's equations (upper panel) and dark Maxwell's equations (lower panel) in the massive dark photon case.
The portal terms play the role of the additional source terms.
If the dark photon is massless ($m_{\gamma'} = 0$), the vector portal ($\varepsilon$) contributions move from dark Maxwell's equations to Maxwell's equations: $\rho \to \big(\rho + \varepsilon \rho'\big)$, $\vJ \to \big(\vJ + \varepsilon \vJ'\big)$ and $\big(\rho' + \varepsilon \rho\big) \to \rho'$, $\big(\vJ' + \varepsilon \vJ\big) \to \vJ'$.
If all portals are closed ($\varepsilon = 0$ and $G_{a\gamma\gamma} = G_{a\gamma\gamma'} = G_{a\gamma'\gamma'} = 0$), the original Maxwell's equations are restored and three particles ($\gamma$, $\gamma'$, $a$) would not communicate with each other.
}
\label{tab:maxwell}
\end{table*}

The kinetic mixing between two Abelian gauge groups provides the vector portal \cite{Holdom:1985ag}
\begin{eqnarray}
\mathcal{L}_\text{kin}  = -\frac{1}{4} \hat{F}_{\mu\nu}\hat{F}^{\mu\nu}+\frac{\varepsilon}{2} \hat{F}_{\mu\nu}\hat{F}'^{\mu\nu}-\frac{1}{4}\hat{F}'_{\mu\nu} \hat{F}'^{\mu\nu}\quad \label{eq:kineticmixing}
\end{eqnarray}
where $\hat{F}$ and $\hat{F}'$ are gauge bosons of the standard model (SM) electromagnetism $U(1)_\text{EM}$ and the dark electromagnetism $U(1)_\text{dark}$, respectively, and a hatted field means that it is not a physical eigenstate yet.
$\varepsilon$ is a dimensionless parameter parametrizing the vector portal.

Depending on the model, the dark photon may get a mass with the Higgs mechanism or Stueckelberg mechanism \cite{Stueckelberg:1900zz}.
It is worth noticing  the physics origin of the light dark photon and light dark fermions.
The small mass can be obtained in various ways \cite{Cleaver:1997nj,Lee:2016ejx,Kim:2017qxo}.
For instance, as proposed in Ref.~\cite{Kim:2017qxo}, chiral representations without gravitational and gauge anomalies can be used.
The only assumption in this new mechanism is the gauge symmetry such as $SU(2) \times U(1)$ in the dark sector, and light dark fermions and light dark gauge boson can be naturally obtained.

Diagonalizing the kinetic mixing and the mass matrix, one can obtain the physical eigenstates by taking the following transformation depending on whether the dark photon is massive or massless \cite{Holdom:1985ag,Galison:1983pa,Dienes:1996zr}.
\bea
\text{ Massive $\gamma'$ case: } \hat A \rightarrow A + \varepsilon A' ~&,& \quad \hat A' \rightarrow A' \label{eq:physical1} \\
\text{ Massless $\gamma'$ case: } \hat A' \to A' + \varepsilon A ~&,& \quad \hat A \rightarrow A \label{eq:physical2}
\eea

The effective Lagrangian for the extended electrodynamics in the presence of the axion and dark photon is
\bea
\begin{split}
\mathcal{L} &= -\frac{1}{4} F F -\frac{1}{4} F' F' + \frac{1}{2} m_{\gamma'}^2 A' A' + \frac{1}{2}\left(\partial_\mu a\right)^2 - \frac{1}{2}m_a^2 a^2 \\
&+ \frac{G_{a\gamma\gamma}}{4} aF\tilde{F} + \frac{G_{a\gamma\gamma'}}{2} aF\tilde{F'} + \frac{G_{a\gamma'\gamma'}}{4} aF'\tilde{F'} + \cal{L}_\text{int}
\end{split}
\label{eq:lagrangian}
\eea
where $J$ ($J'$) is the electromagnetic (dark) current.
\bea
\mathcal{L}_\text{int} =
\begin{cases}
- \big(A + \varepsilon A'\big) J - A' J' & \text{(massive $\gamma'$ case)} \\
- \big(A' + \varepsilon A\big) J' - A J & \text{(massless $\gamma'$ case)} ~~
\end{cases}
\eea
(We do not consider the magnetic or dark magnetic monopoles in this letter.)
In the massive dark photon case, the dark photon can couple to the electromagnetic current with a coupling $\varepsilon e$; in the massless dark photon case, the photon can couple to the dark current with a coupling $\varepsilon e'$.

The axion portal ($G_{a\gamma\gamma}$) as well as the dark axion portal ($G_{a\gamma\gamma'}$, $G_{a\gamma'\gamma'}$) are constructed using the anomaly triangle and the actual couplings depend on the details of the model.
For instance, in the dark KSVZ axion model introduced in Ref.~\cite{Kaneta:2016wvf}, the portal couplings (below the QCD scale) for a massive dark photon are given as
\bea
G_{a\gamma\gamma} &=& \frac{e^2}{8\pi^2} \frac{PQ_\Phi}{f_a} \Big[ 2 N_C Q_\psi^2 - \frac{2}{3} \frac{4+z}{1+z} \Big] , \label{eq:Gagammagamma} \\
G_{a\gamma\gamma'} &\simeq& \frac{e e'}{8\pi^2} \frac{PQ_\Phi}{f_a} \big[ 2 N_C D_\psi Q_\psi \big] + \eps G_{a\gamma\gamma} , \label{eq:Gagammagamma'}\\
G_{a\gamma'\gamma'} &\simeq& \frac{e'^2}{8\pi^2} \frac{PQ_\Phi}{f_a} \big[ 2 N_C D_\psi^2 \big] + 2 \eps G_{a\gamma\gamma'} ,
\eea
where $N_C = 3$ is the color factor and $e$ ($e'$) is the electromagnetic (dark) gauge coupling.
$Q_\psi$ ($D_\psi$) are the electric (dark) charge of the exotic quarks in the anomaly triangle.
$f_a / PQ_\Phi$ is the mass scale of the exotic quarks.
$z \simeq 0.56$ is the mass ratio of the $u$ and $d$ quarks.

The vector portal $\varepsilon$, whose size is constrained to be very small ($\varepsilon^2 \ll 1$) by various data \cite{Essig:2013lka,Alexander:2016aln}, appears in $G_{a\gamma\gamma'}$ and $G_{a\gamma'\gamma'}$.
In the massless dark photon case, the $G_{a\gamma'\gamma'}$ does not contain the $\varepsilon$ term, and the kinetic mixing effect flows in the other direction (i.e., $G_{a\gamma\gamma'} = \cdots + \varepsilon G_{a\gamma'\gamma'}$, $G_{a\gamma\gamma} = \cdots + 2\varepsilon G_{a\gamma\gamma'}$).

Yet there are other terms constructed with dark gauge couplings and independent from the $\varepsilon$, which can be the dominant terms because of the highly constrained $\varepsilon$ parameter.
These terms are nonzero even if $\varepsilon = 0$.

Although we do not consider it in this study, it would be possible to generalize this new portal with a different gauge boson instead of a dark photon, for instance using a dark $Z$ \cite{Davoudiasl:2012ag}, which is a variant of the dark photon with an axial coupling \cite{Davoudiasl:2012qa,Davoudiasl:2013aya,Lee:2013fda,Davoudiasl:2014kua,Davoudiasl:2015bua}) or a light $B-L$ gauge boson (for instance, see Refs.~\cite{Lee:2016ief,Kaneta:2016vkq}).

\begin{table*}[tb]
\begin{tabular}{rcl}
\hline
\\[-6pt]
\( \displaystyle \partial^2 \vE - G_{a\gamma\gamma}\frac{\partial}{\partial t} \Big(\frac{\partial a}{\partial t} \vB+\nabla a \times \vE\Big) - G_{a\gamma\gamma'}\frac{\partial}{\partial t} \Big(\frac{\partial a}{\partial t} \vB'+\nabla a \times \vE'\Big) \)&=&\( 0 \) \\[9pt]
\( \displaystyle \partial^2 \vB + G_{a\gamma\gamma}\nabla \times \Big(\frac{\partial a}{\partial t} \vB+\nabla a \times \vE\Big) + G_{a\gamma\gamma'}\nabla \times \Big(\frac{\partial a}{\partial t} \vB'+\nabla a \times \vE'\Big) \)&=&\( 0 \) \\[9pt]
\hline
\\[-6pt]
\( \displaystyle \big(\partial^2+m^2_{\gamma'}\big) \vE' - G_{a\gamma'\gamma'}\frac{\partial}{\partial t} \Big(\frac{\partial a}{\partial t} \vB'+\nabla a \times \vE'\Big) - G_{a\gamma\gamma'}\frac{\partial}{\partial t} \Big(\frac{\partial a}{\partial t} \vB+\nabla a \times \vE\Big) \)&=&\( 0 \) \\[9pt]
\( \displaystyle \big(\partial^2+m^2_{\gamma'}\big) \vB' + G_{a\gamma'\gamma'}\nabla \times \Big(\frac{\partial a}{\partial t} \vB'+\nabla a \times \vE'\Big) + G_{a\gamma\gamma'}\nabla \times \Big(\frac{\partial a}{\partial t} \vB+\nabla a \times \vE\Big) \)&=&\( 0 \) \\[9pt]
\hline
\\[-6pt]
\( \displaystyle \big(\partial^2+m_a^2\big) a + G_{a\gamma\gamma}\vE \cdot \vB + G_{a\gamma\gamma'} \big(\vE\cdot\vB' + \vE' \cdot\vB\big) + G_{a\gamma'\gamma'}\vE' \cdot \vB' \)&=&\( 0 \) \\[5pt]
\hline
\end{tabular}
\caption{Wave equations for the electromagnetic fields, dark electromagnetic fields, and an axion field when there are no external sources ($\rho = \rho' = 0$ and $J = J' = 0$).
In the massless dark photon case, one can simply take $m_{\gamma'} = 0$.}
\label{tab:oscillations}
\end{table*}

\section{Extended electrodynamics}
Throughout this letter, we take a convention of $\eta_{\mu\nu} = (+,-,-,-)$ with $\partial^2 = \partial_t^2 - \nabla^2$.
The electromagnetic field tensor $F_{\mu\nu} = \partial_\mu A_\nu - \partial_\nu A_\mu$ is then given by
\bea
F_{\mu\nu} =
\begin{bmatrix}
0 & E^1 & E^2 & E^3 \\
-E^1 & 0 & -B^3 & B^2 \\
-E^2 & B^3 & 0 & -B^1 \\
-E^3 & -B^2 & B^1 & 0
\end{bmatrix}
.
\eea

Lagrangian \eqref{eq:lagrangian} provides the equations of motion for the photon, dark photon and axion, respectively, as followings.
\bea
&&\partial_\nu F^{\nu\mu} - G_{a\gamma\gamma} \partial_\nu a \tilde F^{\nu\mu} - G_{a\gamma\gamma'} \partial_\nu a \tilde F'^{\nu\mu} \label {eq:EOM1}
\\
&&~=
\begin{cases}
J^\mu &\text{(massive $\gamma'$ case)} \\
J^\mu + \varepsilon J'^\mu &\text{(massless $\gamma'$ case)}
\end{cases} \nn
\eea
\bea
&&\partial_\nu F'^{\nu\mu} - G_{a\gamma'\gamma'} \partial_\nu a \tilde F'^{\nu\mu} - G_{a\gamma\gamma'} \partial_\nu a \tilde F^{\nu\mu} \label {eq:EOM2}
\\
&&~=
\begin{cases}
- m_{\gamma'}^2 A'^\mu + J'^\mu + \varepsilon J^\mu &\text{(massive $\gamma'$ case)} \\
J'^\mu &\text{(massless $\gamma'$ case)}
\end{cases} \nn
\eea
\bea
\big(\partial^2 + m_a^2\big) a - \frac{G_{a\gamma\gamma}}{4} F \tilde F - \frac{G_{a\gamma\gamma'}}{2} F \tilde F' - \frac{G_{a\gamma'\gamma'}}{4} F' \tilde F' = 0 \nn \\
\label {eq:EOM3}
\eea
Bianchi identities $\partial_\nu \tilde F^{\nu\mu} = 0$ and $\partial_\nu \tilde F'^{\nu\mu} = 0$ are also relevant in deriving the extended Maxwell's and dark Maxwell's equations without source terms.

From the above equations of the motion, one can derive three important sets of the equations that describe the extended electrodynamics in the presence of the axion and dark photon.

\begin{enumerate}
\item[(i)] Extended Maxwell's equations
\item[(ii)] Extended dark Maxwell's equations
\item[(iii)] Wave equations
\end{enumerate}

The extended Maxwell's equations and dark Maxwell's equations are given in Table~\ref{tab:maxwell} for the massive dark photon case; in the massless dark photon case, the vector portal ($\varepsilon$) contributions shift from dark Maxwell's equations to Maxwell's equations (see Table~\ref{tab:maxwell} caption).
As it is quite clear from the equations in the table, the portal terms play a role of the additional sources of the (dark) electromagnetic fields.
For instance, a space-time varying axion field in a magnetic field can produce a dark magnetic field and vice versa.
The wave equations in vacuum (no source) are given in Table~\ref{tab:oscillations}.

\section{\boldmath Photon-Dark photon-Axion oscillations}
\label{sec:oscillations}
One of the striking features of the extended electrodynamics is that in the external magnetic field or dark magnetic field, the light particles (photon, dark photon, axion) can oscillate among themselves.

The oscillation equations among the three particles including all relevant portals can be derived from the equations of motion or Table~\ref{tab:oscillations}.
We present this most general form for the first time.
In the matrix form, they can be rewritten as
\begin{equation}
\big[(\omega^2+\nabla^2)I-\mathcal{M}\big]
\begin{bmatrix} A_{\raisebox{0.5pt}{$\scriptscriptstyle\parallel$}} \\ A'_{\raisebox{0.5pt}{$\scriptscriptstyle\parallel$}} \\ a \end{bmatrix} = 0
\end{equation}
for the parallel direction with the external (dark) magnetic field, with an oscillation frequency $\omega$ of the three fields, and 
\begin{equation}
\mathcal{M}={
\left[ \begin{array}{ccc}
\mathcal{Q}^2+\varepsilon^2 m_{\gamma'}^2 & \varepsilon m_{\gamma'}^2 &- m^2_{a\gamma}\\
\varepsilon m_{\gamma'}^2 & m^2_{\gamma^{\prime}} & -m^2_{a\gamma^{\prime}}\\
-m^2_{a\gamma} &- m^2_{a\gamma^{\prime}} & m^2_a
\end{array}
\right ]} .
\end{equation}
The matrix components vary depending on whether the external field is the magnetic field ($B_T$) or dark magnetic field ($B'_T$) perpendicular to the particle direction (in the transverse direction).
If the dark photon is massless, the direct conversion between the photon and dark photon via the kinetic mixing disappears but is still possible by the axion mediation \cite{Ejlli:2016asd}.

Case (a): in the external magnetic field,
\bea
\mathcal{Q}^2 &=& m^2_\text{plasma}-m^2_\text{QED} \label{eq:Qsquare}\\
m^2_{a\gamma} &=& G_{a\gamma\gamma} \omega B_T \\
m^2_{a\gamma'} &=& G_{a\gamma\gamma'} \omega B_T
\eea

It is also worth to note that the photon may obtain an effective mass in Eq.~\eqref{eq:Qsquare} from the plasma effects and QED effects under the influence of the external magnetic field \cite{Raffelt:1987im}.

For a non-relativistic axion such as the relic cold dark matter axion, the QED mass is negligible compared to plasma mass.
The plasma mass is a function of density, temperature, chemical composition of the medium, and the magnetic field strength.
For example, often we have $m^2_\text{plasma}=\omega_\text{plasma}^2$ near pulsars.
In some special position, this plasma mass can be equal to the axion mass \cite{Huang:2018lxq}.

For a relativistic axion such as the axion emitted from the stellar objects, the QED mass may be relevant.
The QED mass origins  from vacuum polarizability and quantifies photon-photon interactions in the magnetic field.
It is given by $m^2_\text{QED}= 7 \omega^2 \xi$ with $\xi=\frac{e^2}{180\pi^2}\frac{B_T^2}{B_\text{cri}^2}$ ($B_\text{cri}=m_e^2/e\approx4.4 \times 10^{13}$ G), and this is important for high frequency photon compared to axion mass and very strong magnetic field~\cite{Raffelt:1987im}.

Case (b): in the external dark magnetic field,
\bea
\mathcal{Q}^2 &=& m^2_\text{plasma} \\
m^2_{a\gamma} &=& G_{a\gamma\gamma'} \omega B'_T \\
m^2_{a\gamma'} &=& G_{a\gamma'\gamma'} \omega B'_T
\eea

In the external dark magnetic field case, the photon is massless due to the absence of magnetic field.
The dark photon may possibly obtain the dark plasma and dark QED masses depending on the model which might be more relevant in the massless dark photon case.

In the presence of both of the $B$ and $B'$ fields, the mixing feature becomes just a combination of these two cases.
The polarization of the photon (the dark photon) who joins the oscillation corresponds to the parallel direction with $G_{a\gamma\gamma}\omega \vB_T +G_{a\gamma\gamma'}\omega \vB'_T $ $\big(G_{a\gamma\gamma'}\omega \vB_T +G_{a\gamma'\gamma'}\omega \vB'_T\big)$ of which amplitude gives the mixing component $m_{a\gamma}^2$ $(m_{a\gamma'}^2)$.
One can also consider the background electric or dark electric field.
Oscillations in such a case is similar to the magnetic field case with the $B_T$ ($B'_T$) replaced by $E_T$ ($E'_T$), but the polarization is now perpendicular to the electric field or dark electric field.

Because of a rather possibly sizable $G_{a\gamma\gamma'}$, the oscillations here are expected to be greatly enhanced over the results in other models without this term (for instance, see Ref.~\cite{Alvarez:2017eoe}).

\section{Summary and Discussions}
\label{sec:summary}
We extended the effective electrodynamics with contributions from the portals of the axion and dark photon.
We presented the first full expressions for Maxwell's equations, dark Maxwell's equations, and oscillation equations.
Especially, the contribution of the dark axion portal (axion-photon-dark photon vertex) is new and may possibly bring the significant effect as its size is not severely constrained as the other portals.

We note that the additional terms are from the portals that are supposed to connect the SM sector to the mysterious dark sector.
The dark matter or dark sector particles are called dark because we cannot see it, and many search methods for them rely on other types of interactions such as the weak interaction.
The electrodynamics extended with the portals to the dark sector reminds us the dark matter sector may be explored with light.

Possible implications of the extended electrodynamics are huge.
After the extended electrodynamics in the presence of the axion was presented in Ref.~\cite{Sikivie:1983ip}, it was followed by enormous number of works.
This extended electrodynamics using the axion portal ($G_{a\gamma\gamma}$ terms in Maxwell's equations in Table~\ref{tab:maxwell}) is actually used in most axion search experiments \cite{Kim:2008hd}.
We hope our study may also trigger many subsequent works and experiments to probe the dark sector using the electrodynamics.
We list some of the possible directions in the following.

(i) Conversion in the lab: One can consider to design a similar setup as the light-shining-through-wall (LSW) experiments ~\cite{Redondo:2010dp} used for the axion searches.
They apply the external magnetic field to convert an axion to the electric field, using the $G_{a\gamma\gamma} aF\tilde F = -4 G_{a\gamma\gamma} a\vE\cdot\vB$, try to find the resonant signal of the electric field.
One can use $G_{a\gamma\gamma'} aF\tilde F = -2 G_{a\gamma\gamma'} a \big(\vE\cdot\vB' + \vE'\cdot\vB\big)$ to convert the axion into a dark photon under the external magnetic field, and try to find that signal.

(ii) Conversion in the sky:
One can consider the cold dark matter axion in the universe which converts to a photon passing through the dark magnetic field, for instance, an interstellar dark magnetic field, a dark pulsar, a dark magnetic monopole depending on the model \cite{Choi:2018mvk}.
The photon flux density is proportional to the portal coupling $G_{a\gamma\gamma'}$.
This can be detected by future radio telescope experiments such as the Square Kilometer Array (SKA)~\cite{ska:doc} experiment. 
There will be a peak in the radio spectrum corresponding to the axon mass, which also helps to pin down the axion mass.
Due to the new coupling and the possible oscillations among the three light bosons, the radio flux from the neutron star may  be  different from the  benchmark value in Ref.~\cite{Huang:2018lxq}, since it is determined by many sources, such as the property of the mass matrix, the properties of the external (dark) magnetic field, the axion dark matter density and so on.

(iii) Possible extension with a CP-even scalar:
As a possible direction to extend this study, one can consider to introduce a CP-even scalar $\phi$, which may couple to the photon or dark photon as, for instance, $c_A \phi A'_\mu A'^\mu$, $c_B \phi F_{\mu\nu} F^{\mu\nu}$ with a coefficient $c_A$ ($c_B$) of mass dimension $1$ ($-1$).
This could be considered as a scalar boson of the light dark sector, and it may extend Maxwell's equations and wave equations even further.

(iv) Anomaly in the 21cm spectrum:
The 21cm wavelength corresponds to the hyperfine structure in the electron transition of the hydrogen atom.
Recently, an anomalously larger absorption than the expected ($3.8\sigma$ C.L.) was reported by the EDGES Collaboration in their global redshifted 21cm data ($z \approx 17$) \cite{Bowman:2018yin}, which could be explained by the cooling of the hydrogen gas caused by the interaction with the dark matter \cite{Barkana:2018lgd}.
This can be compatible with other astro/cosmological data too if the dark matter has a certain property \cite{Munoz:2018pzp,Fialkov:2018xre,Berlin:2018sjs,DAmico:2018sxd}.
For instance, it was argued to work if a small fraction of the dark matter consists of light particles which couple to the photon with a tiny strength with an additional depletion mechanism for the correct dark matter relic density.
The millicharged dark matter or a fifth force was adopted in Refs.~\cite{Munoz:2018pzp,Fialkov:2018xre,Berlin:2018sjs,DAmico:2018sxd}, but the axion-dark photon system with the extended electrodynamics can naturally fall into this category.
The dark photon (axion) can be a fraction of the dark matter, which can be very light, and it can scatter off the baryons (through $\gamma' N \to a N$ via $G_{a\gamma\gamma'}$ coupling), draining the excess energy from the hydrogen gas, while it is converted into an axion (dark photon).
(This is the so-called `dark Primakoff process' introduced in Ref.~\cite{Kaneta:2017wfh}).
This is qualitatively different from the suggested solutions with a millichargd dark matter or a fifth force in the sense the dark Primakoff process can convert a heavier dark component into a lighter one without an additional depletion mechanism.

Another possibility to explain the 21cm anomaly is to explain the potential bias on the cosmic microwave background (CMB) temperature by the heating of CMB photons \cite{Feng:2018rje}.
There are recent works following this line of approach such as Ref.~\cite{Pospelov:2018kdh} using the axion decay into a pair of dark photons (via $G_{a\gamma\gamma}$) followed by the photon-dark photon oscillation (via $\varepsilon$) and Ref.~\cite{Moroi:2018vci} using the axion to photon oscillation under the primordial magnetic field (via $G_{a\gamma\gamma}$).
This heating of the CMB photons could be accommodated in the extended electrodynamics using $G_{a\gamma\gamma'}$: for example, the $a \to \gamma \gamma'$ decay or the oscillations of the axion to photon in the background dark magnetic field.
Since $G_{a\gamma\gamma'}$ has a dark photon, which does not couple to the SM particles directly, its constraint should be less strict than $G_{a\gamma\gamma}$.
Also, the current data severely constrains $\varepsilon^2 \ll 1$ \cite{Essig:2013lka,Alexander:2016aln}.
Therefore, it is expected that a mechanism using the $G_{a\gamma\gamma'}$ can have a much larger effect than those using the combined effect of the vector portal ($\varepsilon$) and axion portal ($G_{a\gamma\gamma}$).

\vspace{5mm}
\begin{acknowledgments}
This work was supported in part by IBS (Project Code IBS-R018-D1) and NRF Strategic Research Program (NRF-2017R1E1A1A01072736).
We thank S.H.~Yun for conversations.
\end{acknowledgments}



\end{document}